\documentclass[12pt]{article}
\usepackage{jheppub}
\usepackage{mathtools}
\usepackage{mathrsfs}
\usepackage{epstopdf}
\usepackage{verbatim}
\usepackage{slashed}
\def\be{\begin{equation}}
\def\ee{\end{equation}}
\def\bea{\begin{eqnarray}}
\def\eea{\end{eqnarray}}
\def\ba#1\ea{\begin{align}#1\end{align}}
\def\bg#1\eg{\begin{gather}#1\end{gather}}
\def\bm#1\em{\begin{multline}#1\end{multline}}
\def\bmd#1\emd{\begin{multlined}#1\end{multlined}}

\def\({\left(}
\def\){\right)}
\def\[{\left[}
\def\]{\right]}
\def\<{\langle}
\def\>{\rangle}

\begin{document}

\title{Suppressing supersymmetric flavor violations through quenched gaugino-flavor interactions}

\author{James D.\ Wells and Yue Zhao \\
{\it Michigan Center for Theoretical Physics (MCTP) \\ University of
Michigan \\ Ann Arbor, MI 48104}}

\abstract{Realizing that couplings related by supersymmetry (SUSY)
can be disentangled when SUSY is broken, it is suggested that
unwanted flavor and CP violating SUSY couplings may be suppressed
via quenched gaugino-flavor interactions, which may be accomplished
by power-law running of sfermion anomalous dimensions. A simple
theoretical framework to accomplish this is exemplified and the
defeated constraints are tallied. One key implication of the
scenario is the expectation of enhanced top, bottom and tau
production at the LHC, accompanied by large missing energy. Also,
direct detection signals of dark matter may be more challenging to
find than in conventional SUSY scenarios.}

\maketitle

\vfill\eject
%
\section{Introduction}

 Low energy Supersymmetry (SUSY) is
motivated as a solution to the weak-scale hierarchy problem.
However, one of the challenges this theory presents is the potential
introduction of new large flavor and CP violating contributions to
observables that cannot accommodate significant new additions from
new physics.  There are many excellent ideas to solve this problem.
In our work we explore yet another approach to solving the problem
-- the quenching of gaugino-sfermion-fermion interactions. There are
several reasons to investigate this, as will become clearer
throughout the discussion. It is a largely unexplored approach to
solving the flavor problem. It does not add additional finetuning or
naturalness problems compared to other conventional scenarios of
supersymmetry that solve flavor only by fiat. And there may be
interesting new avenues of supersymmetry breaking that lead to this
approach.

Roughly speaking, approaches to supersymmetry have two extremes. One extreme is to think of
supersymmetry as non-existent or broken dramatically, such that there is no SUSY in the low-scale
spectrum that has much hope of being seen anytime soon by current colliders and experiments.
The other extreme is to maintain that supersymmetry is broken very softly, and all superpartners are
``nearby" in the spectrum with full coupling strengths, and are just out of reach of experiment but
could be discovered very soon by slight increases in energy or luminosity at the LHC or by increased
precision on low-energy flavor experiments.

However, nature may choose a middle way, where supersymmetry breaking dynamics is much more
rich than the very simple minded effective approaches we have employed in most studies so far.
Compactifications from higher dimensions, couplings to conformally symmetric sectors, or other so-far
unrecognized dynamics may lead to a supersymmetric spectrum that has apparent unique patterns of
couplings or hard-breaking interactions in the low-scale spectrum. Often times such couplings lead to a form
of supersymmetry that is harder to see at high-energy colliders, but have the advantage of solving some
outstanding problem, such as the flavor and CP violation problems of SUSY.  It is this middle way that we
propose to study, and our specific target is the elimination of gaugino-sfermion-fermion couplings.
Many salient phenomenological features arise by invoking this idea. Furthermore there are reasonable
theoretical approaches that may be able to accomplish exactly this needed pattern. That is our goal.

Any approach that preferentially quenches gaugino-flavor couplings necessarily involves the dynamics and/or
transmission of supersymmetry breaking to the MSSM sector, since it is only through SUSY breaking that
relations between couplings among SM particles can be disentangled from those involving superparticles.
The conventional RG running only modifies the differences between couplings
logarithmically \cite{Cheng:1997sq,Katz:1998br,Kiyoura:1998yt}. So one must do something much more.
One can imagine many different ways to split the couplings by much more than the standard logarithmic
amount, but the approach that we used to illustrative the approach in this study is power-law running.
Fast power-law running can be obtained if some
superparticles are involved in a strongly coupled conformal field
theory (sCFT) and obtain large anomalous dimensions (AD).

In this article, we study the possibility of employing fast power-law
running that differentiates between particles and their
superparticles as an illustrative means by which to quench gaugino-flavor interactions,
thereby eradicating flavor and CP
violation problems in supersymmetry. We show that these problems can
be solved automatically with a reasonable range of power-law
running.  We find in this scenario that the neutralinos are
generically pure states. If the lightest neutralino is the Dark
Matter (DM), the signals in direct detection experiments are
naturally suppressed in the theory.

Neither constraints nor theory require the third generation gaugino-flavor
couplings to be quenched. In the power-law scheme that we consider this means that
 the third generation squarks and slepton
do not need to obtain  large anomalous dimensions like the other two generations.  When
large anomalous dimensions are allowed for first two generations but
not the third, the couplings in different generations become split,
not the squark masses themselves as is the case in other split
family supersymmetry models~\cite{Dimopoulos:1995mi,Pomarol:1995xc}.
Suppressing the first two generation couplings alone can avoid
flavor and CP violation constraints of supersymmetry, and is at
least as beneficial from the naturalness point of view as other
approaches. In addition to describing the theory and detailing the
flavor and CP violation tests that our theory passes, we also
discuss interesting collider physics signatures.


\section{Nelson-Strassler mechanism}

Our goal is to quench gaugino-flavor interactions, and one mechanism
by which this may be possible is through coupling SUSY to a sCFT to
initiate power-law running. Induced power-law running
from a sCFT has been used in the past to address the flavor problem
in related ways. Nelson-Strassler (NS)~\cite{Nelson:2000sn,Nelson:2001mq} was one of
the original approaches to fully exploit this feature. They targeted
the fermion mass hierarchy problem as well as flavor-violating
mixings in the squark/slepton soft SUSY breaking mass matrices and
$A$-terms.  Our illustrative approach targets different parts of the theory, namely
the gaugino-sfermion-fermion interactions, but it
utilizes similar CFT-coupling techniques as NS. Let us first review
NS.

Consider the SM gauge group $S$ times another gauge group $G$. We
assume that $G$ runs to a sCFT at a high scale. Label SM particles
as $X$, which are gauge singlets under $G$. Label exotic particles
as $Z$, which are charged under $G$ and may be charged also under
$S$.

When $G$ runs to its fixed point as a sCFT, ${\rm dim}(X)
> 1$ since $X$ are $G$-singlet operators. One can model build to
give large AD to the first two generations and small ones for the
third generation and $H_u/H_d$. Thus, near the fixed point, the SM
Yukawa coupling $y X_1 X_2 X_3$ is irrelevant for the first two
generations and almost marginal for the third generation. This
induces power law suppression to the Yukawa couplings for the first
two generations while keeping the third ${\cal O}(1)$.


\section{Split Coupling SUSY}

The significant lesson so far is
that large AD over a wide energy range, induced by a sCFT, can
generate large suppression hierarchies from small initial
differences. This is simply due to the anomalous dimension induced
by sCFT, and it does not require SUSY. If the coupling of a SUSY
theory to the CFT does not preserve SUSY, large SUSY breaking
effects can occur as well that might ameliorate the flavor and CP
violation problems.

Let us consider the case where superparticles ($\tilde q$) get large
AD while SM particles ($q$) do not. The fast power law running can
introduce a large hierarchy between two couplings that were
originally related by SUSY. In order to achieve such a scenario, we
couple the MSSM to a sector $G'$ differently for particles than
superparticles. For example, $q$ and $\tilde q$ can couple
differently to particles in $G'$. If $\tilde q$ and $G'$ form a sCFT
at a particular energy scale, $\tilde q$ can obtain large AD while
$q$ does not.

To be more specific, the Lagrangian can be formally written as
\begin{eqnarray}
 \label{formalL}
L\supset \kappa O_{\tilde q} O_{G'}+ a\kappa O_{q} O_{\tilde G'}+...
\end{eqnarray}
Here $...$ includes the source of SUSY breaking. The couplings are
defined schematically such that if $a=1$ supersymmetry is preserved;
i.e., the quarks and squarks both couple to the $G'$ sector in a
manner demanded by supersymmetry invariance. In that case, SUSY is
preserved in the MSSM and supersymmetry breaking is achieved by some
other, perhaps traditional, approach.  This $a=1$ limit reduces to
the NS scenario. In our proposal, we are suggesting a different
limit, $a\ll 1$. This breaks supersymmetry and gives dramatically
different AD to squarks vs.\ quarks.

To be more illustrative, let us consider the squark-gluino-quark
coupling in the Lagrangian. In the MSSM with preserved SUSY, the
coupling between quark and gluon is correlated with the coupling
between squark-gluino-quark,
\begin{eqnarray}
 \label{egVertex}
L\supset -i q^{\dag}\bar{\sigma}^\mu (\partial_\mu-i g_3 A_\mu^aT^a)
q -\sqrt{2}\lambda_3 (\tilde q^* T^a q \tilde g^a+c.c.)
\end{eqnarray}
i.e., $\lambda_3=g_3$. When SUSY is broken, these couplings become
different. In ordinary scenarios, the difference is controlled by a
logarithmic running. However if either the squark or gluino gains
large AD from a sCFT, $\tilde q^* T^a q \tilde g^a$ becomes an
irrelevant operator. Thus $\lambda_3$ enjoys a power law running.
Similarly, the couplings between squarks/sleptons and
electroweakinos can also be suppressed by power law running.

One subtlety we want to emphasize is that the gauge couplings
between superparticles will not be power-law suppressed. This is
guaranteed by gauge symmetry. One can confirm this principle by
absorbing the gauge coupling into the kinetic term of the gauge
boson. The interaction between the gauge boson and a particle is
then directly extracted from the particle's kinetic term constructed
from the gauge-coupling-less covariant derivative. The covariant
derivative maintains itself without alteration upon canonically
normalizing the particle's kinetic term.

\bigskip

The above are the generic aspects that can give rise to our scenario.
For the rest of this section we give some additional comments about possible specific
directions one may wish to pursue to build a specific and complete model. These comments are outside of the mainline of
our present work, and the reader may wish to skip to the next section. However, we find
the richness of realization possibilities encouraging for this scenario and wish to
make a few remarks on them.

To begin with, there are a few examples where a non-SUSY CFT can be
constructed. One is the $\lambda\phi^4$ theory in $(4-\epsilon)$
dimension, which has been used as an example in conformal
sequestering models \cite{Schmaltz:2006qs}. Another class of non-SUSY CFTs
have been conjectured in \cite{Kachru:1998ys}, motivated by the AdS/CFT correspondence.

In contrast to the non-SUSY approaches, one benefit of a SUSY CFT is that the anomalous dimension is
associated with the $R$-charge of field. This makes the theory well
under control computationally. Here we present a model where an approximate SUSY can be
applied in order to control the theory while still generating the
desired feature for Split Coupling SUSY.

Let us embed the MSSM into $N=2$ SUSY content. For example, a quark comes with a
hypermultiplet as $(q,\tilde q, q',\tilde q')$. We label this sector
as $MSSM_2$. There are two ways to split the hypermultiplet in terms
of $N=1$ SUSY. One is $(\{q,\tilde q\}, \{q',\tilde q'\})$ which is
compatible with the SUSY generators of the MSSM. The other way is to
group the hypermultiplet as $(\{q,\tilde q'\}, \{q',\tilde q\})$,
which respects the SUSY generators in $N=2$ also, but whose $N=1$ embedding is orthogonal to
that of the MSSM. In either case, the interactions in $N=2$ are highly
constrained. We explicitly break $N=2$ SUSY in $MSSM_2$ to $N=1$
SUSY by Yukawa couplings in the superpotential. Nevertheless, the $N=2$ SUSY
particle contents still remain.

Now we introduce another sector labeled $G$. Assume $G$ has also
$N=2$ particle content and the interactions within this section
preserve $N=2$ SUSY. Here we emphasize that the $N=2$ SUSY in sector
$G$ is not exact because at least gravity can mediate the $N=2$ SUSY
breaking effects from sector $MSSM_2$ to sector $G$. However, one
generically expects such effects to be small since the running of
dimensionless couplings is logarithmic. And the $N=2$ SUSY is still
approximately preserved.\footnote{Our model is essentially a
dimension deconstructed version of extra dimension model with $N=2$
SUSY being explicitly broken at fixed points of orbifolds.}

Let us assume that there is a relevant operator introducing additional couplings
between $MSSM_2$ and $G$ sectors. However, this relevant operator
only preserves the $N=1$ SUSY which is orthogonal to that of the $MSSM$.
\begin{eqnarray}
 \label{rel}
O_{rel}= M^{n} O_G O_{MSSM_2}
\end{eqnarray}
where $n$ is a positive number which is determined by the dimensions
of $O_G$ and $O_{MSSM_2}$. For example, $O_{MSSM_2}$ could be
identified as a chiral supermultiplet consisting of $\{q',\tilde q\}$,
which preserves the SUSY generator in $N=2$ SUSY orthogonal to the
SUSY generators of the MSSM. Here we emphasize that SUSY is still a good
symmetry at energy scales higher than $M$ because the operator in Eq.
(\ref{rel}) is a relevant operator. From the viewpoint of sector
$G$, when the energy is below $M$, the approximate $N=2$ SUSY is
dramatically broken by this relevant operator, but there is still an
approximate $N=1$ SUSY which is orthogonal to the $N=1$ SUSY of the $MSSM$.
If $(G+O_{rel})$ runs to a strongly coupled CFT at a scale
not far below $M$, the anomalous dimension of $\tilde q$ can be
calculated by its $R$-charge under the approximate $N=1$ SUSY in
sector $G$.

One may worry that although $\tilde q$ obtains a large anomalous
dimension and its coupling to quark and gluino is power law
suppressed, $\tilde q'$ can remain untouched. Especially, if $N=2$
SUSY is preserved in the $MSSM_2$ sector, $\tilde q'-\psi_A-q$ has to
appear, where $\psi_A$ is a fermion paired with a gaugino in the $N=2$
gauge supermultiplet. However, as discussed above, $N=2$ SUSY is not
preserved by interaction terms in $MSSM_2$. Such operators do not
necessarily appear or do not have couplings comparable to the strong gauge
coupling. Further, $\tilde q'$ does not directly couple to our Higgs bosons.
Therefore its mass can be large and does not generate large soft
mass terms to Higgs doublets at 1 loop.


\section{QFT description} Let us describe in more detail the
underlying QFT picture for the suppression and splitting of couplings. We start with a simplified
Lagrangian, writing explicitly only the terms that we are interested
in:
\begin{eqnarray}
 \label{Lag0}
L&\supset &-\frac{1}{4}F^2- |(\partial_\mu-i g A_\mu) \tilde
q_{L/R}|^2-iq_{L/R}^\dag
\slashed{D}  q_{L/R}\nonumber\\
&&-\sqrt{2}\lambda \tilde q_{L/R}^* q_{L/R} \tilde \gamma
-\frac{y}{2}\phi q_L q_R+...
\end{eqnarray}
Here we have kept the squark gauge coupling, the Yukawa coupling
from the superpotential, and the interaction between
squark-gaugino-quark. For simplicity, we consider a $U(1)$ gauge
coupling with charge one. Generalizing to non-abelian gauge symmetry
is straightforward. $\lambda$ is equal to the gauge coupling $g$ if
SUSY is exact.

Now let us integrate out an energy shell and renormalize the
Lagrangian,
\begin{eqnarray}
 \label{Lag1}
L&\supset& -\frac{1}{4}Z_A F^2- Z_{\tilde q_{L/R}}|(\partial_\mu-i
Z_g g A_\mu)  \tilde q_{L/R}|^2\nonumber\\&&-i Z_{q_{L/R}}
q_{L/R}^\dag \slashed{D} q_{L/R} - \sqrt{2}\lambda Z_\lambda \tilde
q_{L/R}^* q_{L/R} \tilde \gamma\nonumber\\&& -\frac{y}{2}Z_y \phi
q_L q_R+...
\end{eqnarray}
Here $Z_A$, $Z_{\tilde q_{L/R}}$ and $Z_{q_{L/R}}$ come from
wavefunction renormalization. $Z_g$, $Z_\lambda$ and $Z_y$ are from
1-PI renormalization of the interaction vertices. If a field obtains
a large AD, its wavefunction renormalization $Z_i$ is power-law
enhanced,
$Z_i\sim(\frac{1}{\epsilon_i})^{2}\sim(M_{c,b}/M_{c,e})^{2\gamma_i}$.
Here $M_{c,b}$($M_{c,e}$) is the beginning (ending) energy scales of
the conformal regime. $\gamma_i$ is the anomalous conformal
dimension.

The 1-PI vertex renormalization factors depend on the details of how
$G'$ couples; it can be very large or $O(1)$. After canonically
normalization,
\begin{eqnarray}
 \label{Lag2}
L&\supset& -\frac{1}{4} F^2- |D_\mu \tilde q_{L/R}|^2-i
q_{L/R}^\dag \slashed{D}\ q_{L/R}\nonumber\\
&&- \sqrt{2}\lambda (Z_\lambda Z_{\tilde q_{L/R}}^{-1/2}
Z_{q_{L/R}}^{-1/2} Z_{\tilde \gamma}^{-1/2})\tilde q_{L/R}^*\
q_{L/R} \tilde \gamma\nonumber\\
&& -\frac{y}{2}Z_y Z_{q_{L}}^{-1/2}Z_{q_{R}}^{-1/2}Z_\phi^{-1/2}\phi
q_L q_R+...
\end{eqnarray}
For gauge coupling terms, the factors of wavefunction
renormalization from matter fields cancel precisely the 1-PI
interaction vertex correction in the covariant derivative terms.
This is required by gauge symmetry. If SUSY is preserved, the
cancelation also happens in the squark-quark-gaugino vertex, i.e.
$Z_\lambda=Z_g\sim (\frac{1}{\epsilon_i})^2$, which guarantees that
the coupling constant $\lambda$ is the same as the gauge coupling
$g$ at all scales.

In contrast, within the NS scenario, cancelation does not occur for
the Yukawa vertex. Appealing to the non-renormalization theorem,
$Z_y=1$ is fixed. Nevertheless, wavefunction renormalization
introduces a power-law suppression on the Yukawa couplings as
$y_q\sim \epsilon_{q_L}\epsilon_{q_R}$.

\begin{figure}[t]
\begin{center}
\includegraphics[width=0.40\textwidth]{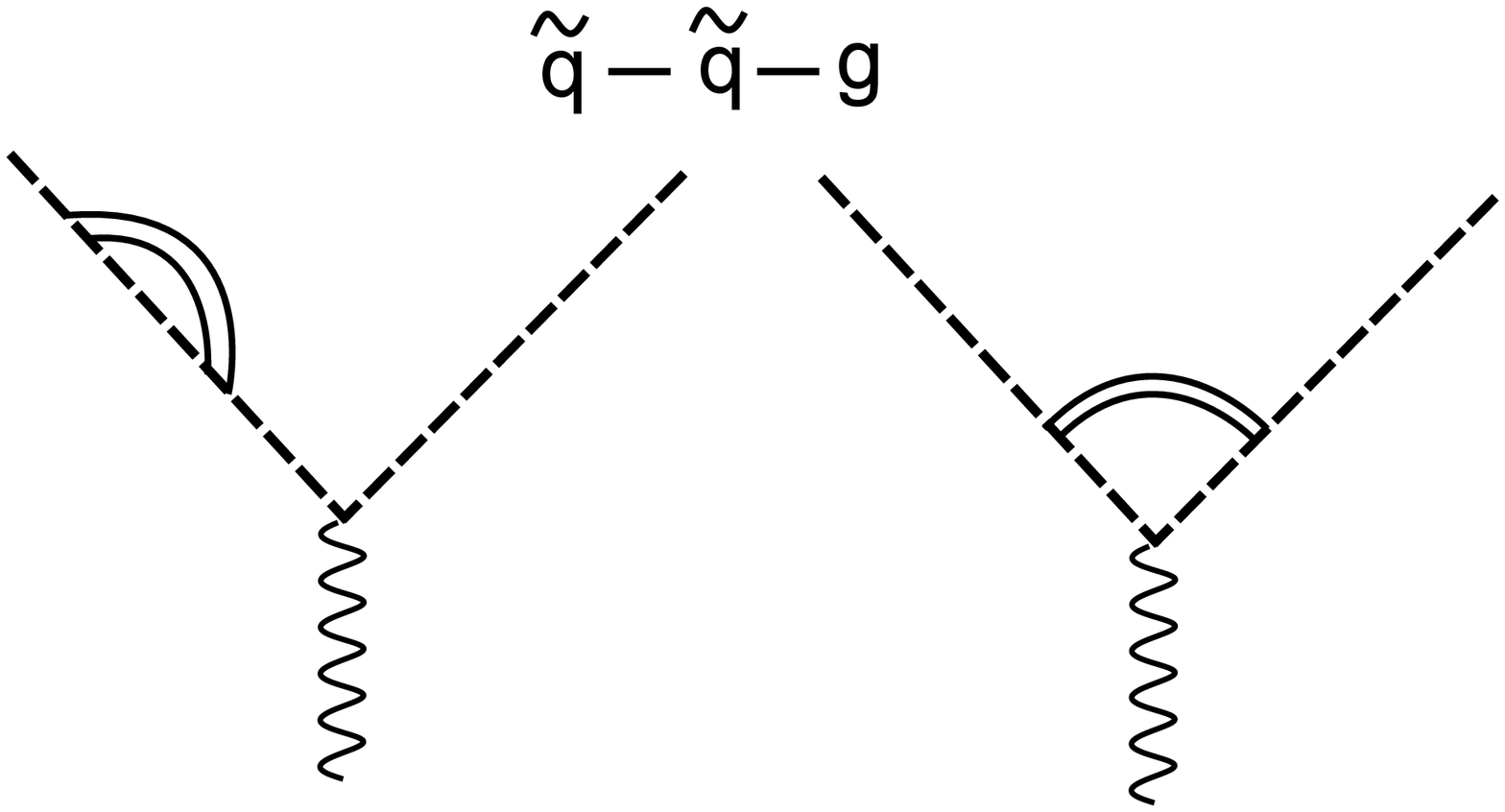}\ \ \ \ \
\includegraphics[width=0.17\textwidth]{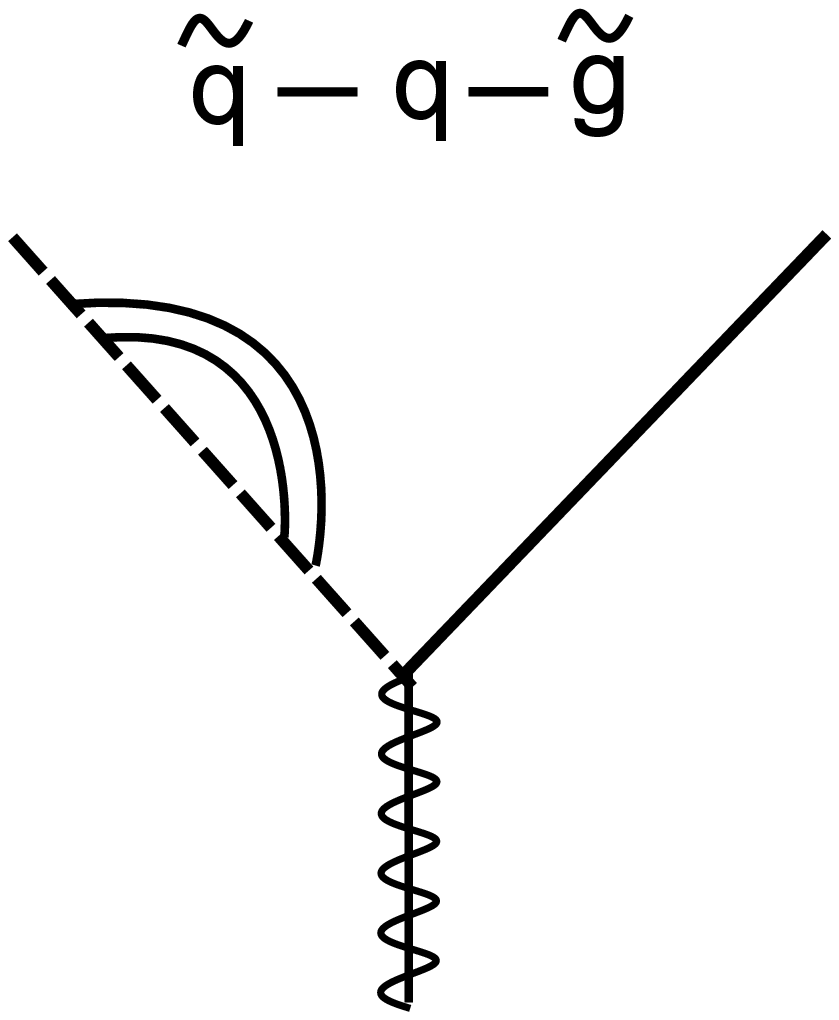}
\hspace*{0.35cm} \caption{Here we show a few examples of Feynman
diagrams where a strongly coupled CFT may contribute at 1-loop. Double
lines indicate large renormalization from the strongly coupled CFT.  If
sCFT only couples to squarks, both the squark propagator and its gauge
coupling vertex receive large corrections, as shown in the first two
diagrams. However, there is no large 1-PI vertex correction to
$\tilde q-q-\tilde g$ vertex as shown in the third diagram. }
\label{fig:UV_FP1}
\end{center}
\end{figure}

Now let us turn to our scenario where the squark and quark couple
differently to the sector $G'$. If squark+$G'$ runs to a sCFT at a
particular scale, the squark may obtain a large AD, i.e. $Z_{\tilde
q}\sim(1/\epsilon_{\tilde q})^2\gg 1$. Since SUSY is broken,
$Z_\lambda$ no longer has a rigid supersymmetric relation with
$Z_g$. This disconnect is maximal if $G'$ does not directly couple
to quarks and gauginos, but does couple to squarks. This results in
$Z_\lambda$, $Z_q$ and $Z_{\tilde \gamma}$ expected to be ${\cal
O}(1)$, whereas $Z_{\tilde q_{L/R}}^{-1/2}\sim \epsilon_{\tilde
q_{L/R}}$. Thus, the only significant suppression factors in our
theory arise from wave function renormalization factors on squarks
and sleptons, and possibly higgsinos also as will be discussed
later. This leads to quenching of the gaugino-flavor interactions
and possibly also the higgsino-flavor interactions. The mismatch
between couplings of quarks/leptons vs.\ squarks/sleptons is
schematically illustrated in Fig.~\ref{fig:UV_FP1}.

The considerations outlined in this section enable us to define various options of spectra for
the masses of the minimal supersymmetric particles. Various benchmark possibilities will be
discussed in detail later in sec.\ \ref{sec:benchmark}.

\section{Soft SUSY breaking terms}

It is also important to
understand how the soft mass terms run during the conformal region.
Naively, one would expect that the soft mass terms are large in our
scenario because the SUSY relations among couplings are broken at
the beginning of the conformal region, i.e. $E_{c,b}$. Especially,
one would generically expect that the squark soft mass square is
quadratically sensitive to $E_{c,b}$.  However, the running of soft
terms is quite subtle in the conformal region and highly depends on
how to couple the  $MSSM$ to the sCFT. It turns out that such running can
be estimated if the CFT has an approximate supersymmetry. This was studied in
detail in \cite{Nelson:2001mq}.

Let us take
our example where the $MSSM$ is embedded into an $N=2$ SUSY framework.
In such a model, from the viewpoint of sector $G$, SUSY is still an
approximate symmetry and the running of soft terms can still be
estimated in the context of SUSY. In \cite{Nelson:2001mq}, the authors
show that some particular linear combinations of soft masses can be
suppressed during RG flow in the conformal region, while others are not,
depending on the residual global symmetries when coupling the two
sectors. Particularly, if there is an exact global symmetry
preserved in the Lagrangian when coupling to the sCFT, there is a
particular linear combination of soft mass squares unaffected by RG
flow.

On the other hand, flowing into the IR after the conformal region,
the soft mass terms run as normal. Since the SUSY-related couplings
do not match anymore, one expects the soft mass squares to be at
least quadratically sensitive to the end of conformal region, i.e.\
$E_{c,e}$. As we will discuss later, the third generation squarks do
not necessarily obtain large AD in order to avoid flavor/CP
problems. It is possible that only the first two generation squarks
get large AD while the 3rd generation does not. This can be helpful
from the fine-tuning point of view because corrections to the Higgs
soft mass square, induced by the large top Yukawa coupling, depend
quadratically on the stop mass but not $E_{c,b}$. In this article,
we do not specify a UV model for the sCFT, and therefore we treat
these soft masses as free parameters. Also, we will mainly focus on
the scenario where the 3rd generation is treated differently from
the first two generations.

\section{Remedying the SUSY flavor/CP problem}

The primary motivation of our
work is to suppress the flavor and CP problems of supersymmetry via
quenched gaugino-flavor couplings which can be accomplished, for example,
by power-law suppressions of squark/slepton couplings.
We first review the constraints and then apply those constraints to
our scenario and show that reasonable parameters lead to safe
phenomenology.

Flavor and CP measurements put stringent constraints on the
parameter space in supersymmetry \cite{Gabbiani:1996hi}. All these
constraints can be characterized by flavor off-diagonal soft SUSY
breaking mass terms.

Let us first focus on hadronic systems. There are two classes of
processes, characterized by the change of flavor number, i.e.
$\Delta F=1$ or 2.

$\Delta F=2$ processes can be described by dimension 6 operators,
e.g.
\begin{eqnarray}
 \label{F2operator}
O=\frac{1}{\Lambda^2}(\bar d_L\gamma^\mu s_L)(\bar d_L\gamma_\mu
s_L).
\end{eqnarray}
These are induced by a box diagram with four squark-gluino-quark
vertices. Since the flavor number is changed by 2, we need at least
two insertions of flavor off-diagonal soft mass elements. Thus the
suppression scale of such operators scales as
\begin{eqnarray}
 \label{F2operatorScale}
\frac{1}{\Lambda^2}\sim \lambda^4
\frac{(\delta^a_{ij})_{AB}^2}{m_{SUSY}^2}
\end{eqnarray}
$(\delta^a_{ij})_{AB}=\frac{(m^a_{ij})^2_{AB}}{m_{SUSY}^2}$,
assuming all flavor-diagonal squark masses and gluino mass equal to
$m_{SUSY}$ for simplicity. $\{A,B\}$ label left/right squarks,
$\{a\}$ indicates up/down type, $\{i,j\}$ are generation indices. We
also explicitly write the dependence on squark-gluino-quark
couplings ($\lambda$), without specifying flavors -- flavor
violations are all embedded in the $\delta_{ij}$ factors. If
$\lambda$ is much smaller than the gauge coupling $g_3$, which
power-law running can accomplish ($\lambda\sim \epsilon_{\tilde
q}g_3$), the constraints from flavor/CP can be weakened.

Assuming flavor universality, the strongest constraints on $\Delta
F=2$ processes are from from $K-\bar K$ mixing. This system also
manifests a CP-violating phenomenon characterized by the parameter
$\varepsilon$. We suppress all indices of $\delta$ assuming all
$(\delta^a_{ij})_{AB}$ are comparable and complex. From $K-\bar K$
system measurements, constraints become
\begin{eqnarray}
 \label{KKbarMixing}
\left(\frac{\lambda}{g_3}\right)^2 |\textrm{Re}(\delta)| &<&10^{-3}
\left(\frac{m_{SUSY}}{500\
\textrm{GeV}}\right)\nonumber\\
\left(\frac{\lambda}{g_3}\right)^2 |\textrm{Im}(\delta)| &<&10^{-4}
\left(\frac{m_{SUSY}}{500\ \textrm{GeV}}\right).
\end{eqnarray}
If $\frac{\lambda}{g_3}$ is smaller than $10^{-2}$, which is easily
accomplished by our power-law running scenario, one can have ${\cal
O}(1)$ flavor mixing with TeV scale squarks.

For $\Delta F=1$ processes, both box diagrams and penguin diagrams
contribute. For illustration, we show one operator for each kind:
\begin{eqnarray}
 \label{F1operators}
O_{box}=\frac{(\bar d_L\gamma^\mu s_L)(\bar q_L\gamma_\mu
q_L)}{\Lambda_{\rm box}^2}; O_{pen}=\frac{\bar d_L\sigma^{\mu\nu}s_R
F_{\mu\nu}}{\Lambda_{\rm pen}}.
\end{eqnarray}
Since $\Delta F=1$, one only needs one flavor-changing mass
insertion in the loop. Thus we have
\begin{eqnarray}
 \label{F1operatorScale}
\frac{1}{\Lambda_{box}^2}\sim \lambda^4
\frac{(\delta^a_{ij})_{AB}}{m_{SUSY}^2}.
\end{eqnarray}
Penguin diagrams have subtlety on chirality. Depending on whether
the chirality is changed in the soft mass insertion, the operator
effectively is either dimension 5 or 6:
\begin{eqnarray}
 \label{F1operatorScale}
\frac{1}{\Lambda_{pen}}\sim c_1\lambda^2
\frac{(\delta^a_{ij})_{AA}m_q}{m_{SUSY}^2}+c_2\lambda^2
\frac{(\delta^a_{ij})_{LR}}{m_{SUSY}}.
\end{eqnarray}

The strongest constraints on $\Delta F=1$ processes come from a
CP-violating measurement in Kaon system, $\varepsilon'/\varepsilon$.
If the chirality is not changed by a soft mass insertion, i.e.\ by
$(\delta^a_{ij})_{AA}$, the box and penguin contributions tend to
cancel with each other when $\lambda=g_3$. However the box diagram
has a different $\lambda$ dependence from that of the penguin
diagram, and when $\lambda\ll g_3$ the penguin diagrams always
dominate.

Assume all $\delta$'s are comparable complex numbers,we have
\begin{eqnarray}
 \label{DeltaS1}
\left(\frac{\lambda}{g_3}\right)^2 |\textrm{Im}(\delta)_{AA}|
&<&10^{-1} \left(\frac{m_{SUSY}}{500\
\textrm{GeV}}\right)^2\nonumber\\
\left(\frac{\lambda}{g_3}\right)^2  |\textrm{Im}(\delta)_{LR}|
&<&10^{-5} \left(\frac{m_{SUSY}}{500\ \textrm{GeV}}\right).
\end{eqnarray}
When $m_{SUSY}$ is ${\cal O}({\rm TeV})$, if
$\frac{\lambda}{g_3}<3\times10^{-3}$, $\delta$ can be ${\cal O}(1)$.

Flavor-changing lepton decays, $\ell_i\to \ell_j+\gamma$, especially
muon decay, impose constraints on the leptonic sector. Assuming
$m_{\tilde l}\sim m_{\tilde \gamma}\sim m_{SUSY}$,
\begin{eqnarray}
 \label{MuonDecay}
\left(\frac{\lambda'}{e}\right)^2 |\textrm{Im}(\delta)_{AA}|
&<&10^{-2} \left(\frac{m_{SUSY}}{100\
\textrm{GeV}}\right)^2\nonumber\\
\left(\frac{\lambda'}{e}\right)^2 |\textrm{Im}(\delta)_{LR}|
&<&10^{-6} \left(\frac{m_{SUSY}}{100\ \textrm{GeV}}\right).
\end{eqnarray}
$\lambda'$ is the slepton-photino-lepton coupling. If
$\frac{\lambda'}{e}<10^{-3}$, one can get around the muon decay
constraint with ${\cal O}(1)$ mixing in the slepton soft mass
matrix, even if the slepton and photino are as light as 100 GeV.

Finally, the flavor diagonal soft mass matrix, i.e.
$(\delta_{ii})_{LR}$, can be strongly constrained by radiative mass
corrections and electric dipole moments.

The radiative mass corrections scale as
\begin{eqnarray}
 \label{RadMass}
\Delta m_q\sim \lambda^2 m_{SUSY} \textrm{Re}
(\delta^q_{ii})_{LR}\nonumber\\ \Delta m_l\sim \lambda'^2 m_{SUSY}
\textrm{Re} (\delta^l_{ii})_{LR}.
\end{eqnarray}
$m_{SUSY}$ refers to the gluino and photino masses for the quark and
lepton mass corrections respectively. The strongest constraint comes
from the first generation. Requiring the radiative masses to not
exceed the quark/lepton mass, one gets
\begin{eqnarray}
 \label{RadMassConstraints}
\left(\frac{\lambda}{g_3}\right)^2  \textrm{Re} (\delta^q_{11})_{LR}
< 2\times 10^{-3}
\left(\frac{500\ \textrm{GeV}}{m_{SUSY}}\right)\nonumber\\
\left(\frac{\lambda'}{e}\right)^2  \textrm{Re} (\delta^l_{11})_{LR}
< 8\times 10^{-3} \left(\frac{100\ \textrm{GeV}}{m_{SUSY}}\right).
\end{eqnarray}

The electric dipole moments of the neutron and electron scale as
\begin{eqnarray}
 \label{Dipole}
\frac{d_q}{e}\sim\lambda^2\frac{\textrm{Im}(\delta^q_{11})_{LR}}{m_{SUSY}};\
\ \
\frac{d_e}{e}\sim\lambda'^2\frac{\textrm{Im}(\delta^l_{11})_{LR}}{m_{SUSY}}.
\end{eqnarray}
The constraints are,
\begin{eqnarray}
 \label{DipoleConstraints}
\left(\frac{\lambda}{g_3}\right)^2\textrm{Im}(\delta^q_{11})_{LR}<3\times 10^{-6}\left(\frac{m_{SUSY}}{500\ \textrm{GeV}}\right)\nonumber\\
\left(\frac{\lambda'}{e}\right)^2\textrm{Im}(\delta^l_{11})_{LR}<4\times
10^{-7}\left(\frac{m_{SUSY}}{100\ \textrm{GeV}}\right).
\end{eqnarray}
If $\delta$'s are ${\cal O}(1)$ complex numbers, one needs
$(\frac{\lambda}{g_3})$ or $(\frac{\lambda'}{e})$ smaller than
${\cal O}(10^{-3})$ for light squarks and sleptons, which are
readible accessible with power-law running suppressions.

\section{The first two generations versus the third} The strongest
constraints on flavor physics are transitions among the light
fermions. In our baseline theory, only the first two generations of
squarks/sleptons obtain large AD, while the third generation does
not\footnote{We envision giving the same AD to all superparticles
within each generation in order to avoid quadratic sensitivities
from $D$-terms.}.

The benefit of doing this is to maintain naturalness. As discussed
previously, the running of soft mass terms depends on the UV model
controlling the conformal region. However after exiting the
conformal region, i.e.\ below $E_{c,e}$, the sCFT no longer helps
suppress soft terms. If the stop were to obtain a large AD it would
couple differently to the Higgs boson. Thus, fine tuning would then
be quadratically related to $E_{c,e}$. This would be similar to
(mini) split SUSY
\cite{Wells:2003tf,ArkaniHamed:2004fb,ArkaniHamed:2004yi,Wells:2004di,Giudice:2004tc,Arvanitaki:2012ps,ArkaniHamed:2012gw}.

In contrast, for our choice, the naturalness problem is much less
severe since only the first two generations get large AD. In this
case, in order to maintain $m_{H_u}\sim$ TeV naturally, $E_{c,e}$
needs to be smaller than PeV. The soft mass of squarks in the first
two generations are
generically heavy 
${\cal O}(E_{c,e}^2/16\pi^2)$ which further helps to avoid flavor/CP
problems\footnote{We thank Stuart Raby for pointing out this
feature.}.

One interesting phenomenological consequence is its unique collider
signatures. First, the production of superparticles are not
dramatically suppressed because all squarks and the gluino have
unsuppressed gauge couplings. The production channels of
squark/gluino are mainly through gluon fusion and $q\bar q\to g^*$,
whereas quark-gluino associated production is suppressed.

We assume the LSP is a neutralino which has suppressed couplings to
the first two generations of squarks. We also assume all squarks
have comparable masses for simplicity. When the gluino is heavier,
the squark directly decays to neutralino. Interestingly the dominant
decay channel would be through mixing to the third generation
squarks, $\tilde u\to\tilde t\to t\tilde \chi^0$, where $\tilde
\chi^0$ is a
neutralino. 
Thus the signature would be mainly two third-generation quarks plus
missing energy (MET). If two gluinos are produced, the signature
would be four bottom/stop quarks plus MET.

If squarks are heavier, gluino production is dominant. They will
decay through off-shell squarks that will in turn produce four
third-generation quarks plus MET. If squarks are produced, there can
be six third-generation quarks plus MET. However, it is possible
that the stop/sbottom-gluino-top/bottom vertices receive a stronger
suppression than those with neutralino, squarks then would decay to
neutralino without passing through a gluino. The detailed signatures
depend on the UV model. Nevertheless, the main theme is clear: the
preponderance of third-generation quarks and leptons accompanied by
missing energy.

\section{Dark matter and direct detection}

In the MSSM, the
lightest neutralino is a good candidate for DM.  At tree level, they
can interact with nucleons through $t$-channel higgs exchange,
$t$-channel $Z$ exchange and $s$-channel squark exchange, which
potentially allows for their discovery in suitable laboratory
detectors.

The mixture among the neutralino interaction eigenstates to form
mass eigenstates originates from their couplings to higgsinos, i.e.
$h^\dag\tilde h \tilde b$ and $h^\dag\tilde h \tilde w$. If the
mixing is small, a neutralino barely couples to the Higgs and $Z$
bosons. If either higgsinos or the gauginos (bino/wino) obtain large
AD, the mixing can be highly suppressed. Thus it is quite generic to
expect the lightest neutralino to be almost a pure state. If the LSP
is a higgsino and all neutralinos have mass $\cal{O}(\textrm{TeV})$,
the suppression factor needs to be no stronger than ${\cal
O}(10^{-2})$, in order for the higgsino to not be a Dirac fermion
from the DM detection point of view. However, a pure state like this
creates experimental detection challenges. In particular, the mixing
is small even when the bino, wino and higgsino have comparable
masses, and so DM detection via Higgs or $Z$ exchange is not
effective method to problem the theory.

$s$-channel squark exchange may also induce scattering between DM
and nucleons. However, a strongly coupled CFT can suppress such
couplings efficiently. The conclusion remains that the strong
constraints on SUSY from lack of direct detection of DM do not apply
here.

\section{Coupling hierarchies and benchmark scenarios}
\label{sec:benchmark}

Let us briefly summarize the physics effects introduced by the split
coupling scenario. For simplicity, we assume the strongly coupled
sector does not introduce large 1-PI vertex renormalization except
for gauge vertices. After canonical normalization, power-law
suppression factors to non-gauge vertices arise from the external
leg corrections, which include squarks, sleptons and Higgsinos. We
identify these factors $\epsilon_{\tilde q}$, $\epsilon_{\tilde
\ell}$, and $\epsilon_{\tilde h}$, all of which are $\epsilon \ll
1$.  However, we will assume that third generation squarks and
sleptons do not receive large power-law suppression factors,
$\epsilon_{\tilde t,\tilde b,\tilde \tau}\sim 1$. This choice is not
necessary for the viability and unique attractiveness of the theory;
however, it is allowed experimentally due to the relatively weak
flavor bounds on third generation processes and does not exacerbate the
naturalness problem.

Here we write a few examples of how the vertices are altered:
\begin{eqnarray}
 \label{eq:vertices}
h^2\tilde \ell^2\rightarrow (\epsilon_{\tilde \ell}^2) h^2 \tilde
\ell^2;\  \tilde q \tilde g q\rightarrow (\epsilon_{\tilde q})\tilde
q \tilde g q; \ \ h\tilde h\tilde w\rightarrow(\epsilon_{\tilde
h})h\tilde h\tilde w,
\end{eqnarray}
where we have ignored Lorentz indices, etc.
Other interactions are not suppressed in our approach, most notably
$h\bar f f$ (where $f=q,\ell,\nu$) and all gauge interactions
arising from covariant derivative interactions, such as $A_\mu \bar
f\gamma^\mu f$, $A_\mu \tilde f^*\partial^\mu \tilde f$, etc. Furthermore, as
we have discussed earlier, one is likely to be able to build the soft supersymmetric
masses to obtain any value wanted. For this reason, we view them as unconstrained
parameters from the theory point of view, which enables one freely to pursue spectra
that minimize the finetuning of electroweak symmetry breaking, for example, and more
importantly suggests that all mass scales and hierarchies should be considered when
constructing benchmark models and
searching for experimental signatures of the scenario.

\section{Discussion}

In this article, we have discussed the
possibility of quenching gaugino-flavor and higgsino-flavor interactions for the
purposes of remedying the flavor problem of SUSY.  There are potentially many
different ways to achieve this phenomenological aim. For illustrative purposes
we have considered the prospect of coupling the MSSM to a
SUSY-breaking sector $G'$ which flows to a sCFT at a particular
energy scale. The SUSY flavor/CP problems are naturally solved by quenching
the gaugino-flavor couplings. Such split coupling models are analogues
to split SUSY or split family models, depending
on whether the third generation squarks also get large anomalous dimensions.
However, instead of having a split mass spectrum, it is the SUSY-related
couplings that develop large separations. Neutralino DM is likely to
be a pure state, whose direct detection signals are suppressed.

If only the first two generations of squarks get large
anomalous dimensions, the
collider signatures are similar to those in the natural SUSY
approach, where third generation quarks and leptons dominate the
final state, but with different signal rates. As for the SM fermion
mass hierarchy, it can be solved separately by applying the
NS-mechanism. In that case, the superparticle couplings would be
further suppressed. Finally, it is also interesting to consider the
suppression of $R$-Parity violating vertices through a similar
mechanism. We leave the details of this possibility for future
study.

\section*{Acknowledgments} We thank S.\ Martin and A.\ Pierce for
helpful discussions. This work is supported in part by the U.S. Dept
of Energy under grant DE-SC0007859.

\end{document}